\documentclass[fleqn,10pt]{wlscirep}
\usepackage[utf8]{inputenc}
\usepackage[T1]{fontenc}
\usepackage{here}
\title{Estimating survival probability using the terrestrial extinction history for the search for extraterrestrial life}

\author[1,2,*]{Kohji Tsumura}

\affil[1]{Department of Natural Science, Faculty of Science and Engineering, Tokyo City University, Setagaya, Tokyo, 158-8557, Japan}
\affil[2]{Frontier Research Institute for Interdisciplinary Sciences, Tohoku University, Sendai, Miyagi, 980-8578, Japan}

\affil[*]{ktsumura@tcu.ac.jp}

\begin{abstract}
Several exoplanets have been discovered to date, and the next step is the search for extraterrestrial life.
However, it is difficult to estimate the number of life-bearing exoplanets because our only template is based on life on Earth.
In this paper, a new approach is introduced to estimate the probability that life on Earth has survived from birth to the present based on its terrestrial extinction history.
A histogram of the extinction intensity during the Phanerozoic Eon is modeled effectively with a log-normal function,
supporting the idea that terrestrial extinction is a random multiplicative process.
Assuming that the fitted function is a probability density function of extinction intensity per unit time, 
the estimated survival probability of life on Earth is $\sim 0.15$ from the beginning of life to the present. 
This value can be a constraint on $f_i$ in the Drake equation, which contributes to estimating the number of life-bearing exoplanets.
\end{abstract}

\begin{document}
\flushbottom
\maketitle

\thispagestyle{empty}

\section*{Introduction}

Since the first discovery of the exoplanets in the 1990s \cite{1992Natur.355..145W, 1995Natur.378..355M}, 
expectations for the discovery of extraterrestrial life have increased.  
The most likely outcomes are the discovery of traces of life in our Solar System such as Mars, Europa or Enceladus via in-situ explorations made by spacecrafts, 
or the discovery of biosignatures on exoplanets in their host star's habitable zones via astronomical high-resolution spectroscopic observations.
On the other hand, there is a slight possibility that an advanced civilization will be found before these discoveries are made. 
For example, the Square Kilometer Array (SKA), which is the international radio telescope project currently under development, 
will be capable of detecting leakage emissions from Earth-level civilizations within 100 pc \cite{2015aska.confE.116S}. 
Recently, the Five-hundred-meter Aperture Spherical radio Telescope (FAST) in China has been conducting Search for Extraterrestrial Intelligence (SETI) observations \cite{Zhang2020}.
Therefore, estimating the number of planets with extraterrestrial life in our Milky Way galaxy is important. 

One way to make this estimate is to use the Drake equation,
which is the famous algebraic expression for quantifying the number of communicative civilizations in our Galaxy \cite{1961PhT....14...40D}. 
The Drake equation is generally expressed as \cite{1963PSS...11..485S}:
\begin{equation}
N = R_{\ast} \cdot f_{p} \cdot n_{e} \cdot f_{l} \cdot f_{i} \cdot f_{c} \cdot L
\end{equation}
\begin{quote}
\begin{description}
  \item[$N$:] The number of civilizations in our Galaxy with which communication might be possible.
  \item[$R_{\ast}$:] The mean rate of star formation averaged over the lifetime of the Galaxy.
  \item[$f_{p}$:] The fraction of stars with planetary systems.
  \item[$n_{e}$:] The mean number of planets in each planetary system with environments favorable for the origin of life.
  \item[$f_{l}$:] The fraction of such favorable planets on which life does develop.
  \item[$f_{i}$:] The fraction of such inhabited planets on which intelligent life with manipulative abilities arises during the lifetime of their local sun.
  \item[$f_{c}$:] The fraction of planets populated by intelligent beings on which an advanced technical civilization arises during the host star's lifetime.
  \item[$L$:] The lifetime of the technical civilization.
\end{description}
\end{quote}
There are reliable estimates for the first three factors ($R_{\ast} \cdot f_{p} \cdot n_{e} \sim 0.1$\cite{10.1093/mnras/staa512})
based on recent astronomical observations of exoplanets, protoplanetary disks, and star-forming regions. 
However, because we have not yet discovered any extraterrestrial life, nor elucidated the origins of terrestrial life, 
the other remaining factors are highly conjectural owing to the one-sample statistics of Earth.

We discuss $f_{i}$ among these conjectural factors in the Drake equation in this paper.
Previous estimates of $f_{i}$ range from pessimistic ($f_{i} \sim 0$) to optimistic ($f_{i} \sim 1$). 
In general, while many physicists and astronomers prefer the optimistic value, 
many biologists prefer a value several orders of magnitude smaller \cite{1980QJRAS..21..267T, Lineweaver2008}.
Table \ref{tab:fi} shows the various estimated values of $f_{i}$ and $f_{l} \cdot f_{i} \cdot f_{c}$ to date.

\begin{table*}[ht]
\centering
\caption{Previous estimates of $f_{i}$ and $f_{l} \cdot f_{i} \cdot f_{c}$}
\begin{tabular}{|c|c|c|}
\hline
Estimated value of $f_{i}$ & Estimated value of $f_{l} \cdot f_{i} \cdot f_{c}$ & Reference \\
\hline
$\sim 0.1$ & $\sim 0.01$ & \cite{1963PSS...11..485S} \\
$\sim 1$ & 0.1 - 0.2 & \cite{1963icse.book.....C} \\
$\sim 1$ & $\sim 0.5$ & \cite{1963ST....26..258C} \\
$\sim 1$ & $\sim 1$ & \cite{1965cae..book..323D} \\
$\sim 1$ & $ > 0.1$ & \cite{1975Icar...25..360O} \\
$\sim 1$ & $ 0.01$ & \cite{1975Icar...25..368F} \\
--- & $< 10^{-10}$ & \cite{1980QJRAS..21..267T} \\
--- & 0.01 & \cite{1981QJRAS..22..380W} \\
0.01 & $10^{-4}$ & \cite{1992iaot.book.....D} \\
0.01-0.1 & --- & \cite{2009IJAsB...8..121F} \\
0.2 & 0.02 & \cite{2010AcAau..67.1366M} \\
--- & $> 1.7 \cdot 10^{-11}$ & \cite{2016AsBio..16..359F} \\
0.5 & 0.05 & \cite{2019AcAau.155..118B} \\
$\sim 1$ & $< 10^{-40}$ & \cite{Totani2020} \\
$\sim 0.15$ & --- & this work \\
\hline
\end{tabular}
\label{tab:fi}
\end{table*}

This paper introduce a new approach to estimating the probability that life on Earth has not gone extinct since the birth of life, $f_{i, \oplus}$.
Since its birth, life on Earth has gone through many extinction events due to various random external factors, such as changes in the environment or impacts of meteorites. 
Extinction events of high intensity (where a significant fraction of species disappear) occur much less frequently than events of low intensity.
The fossil record in the Phanerozoic Eon, which covers 540 Myr to the present \cite{Sepkoski2002, 2005Natur.434..208R},
indicates that a histogram of extinction intensity can be well modeled by a log-normal distribution.
This log-normal distribution of extinction was converted into a cumulative probability that life on Earth survives up to the present, $f_{i, \oplus}$, 
by ``continuing to win the lottery of extinction'' since its birth.
The obtained survival probability, $f_{i, \oplus}$, can be a template for estimating $f_{i}$ in the Drake equation, 
or other factors for estimating the number of life-bearing exoplanets,
assuming that life on any other exoplanets essentially always becomes complex if it does not become extinct first.

\section*{Histogram of the Terrestrial Extinction History}

\begin{figure}[H]
\centering
\includegraphics[scale=0.3]{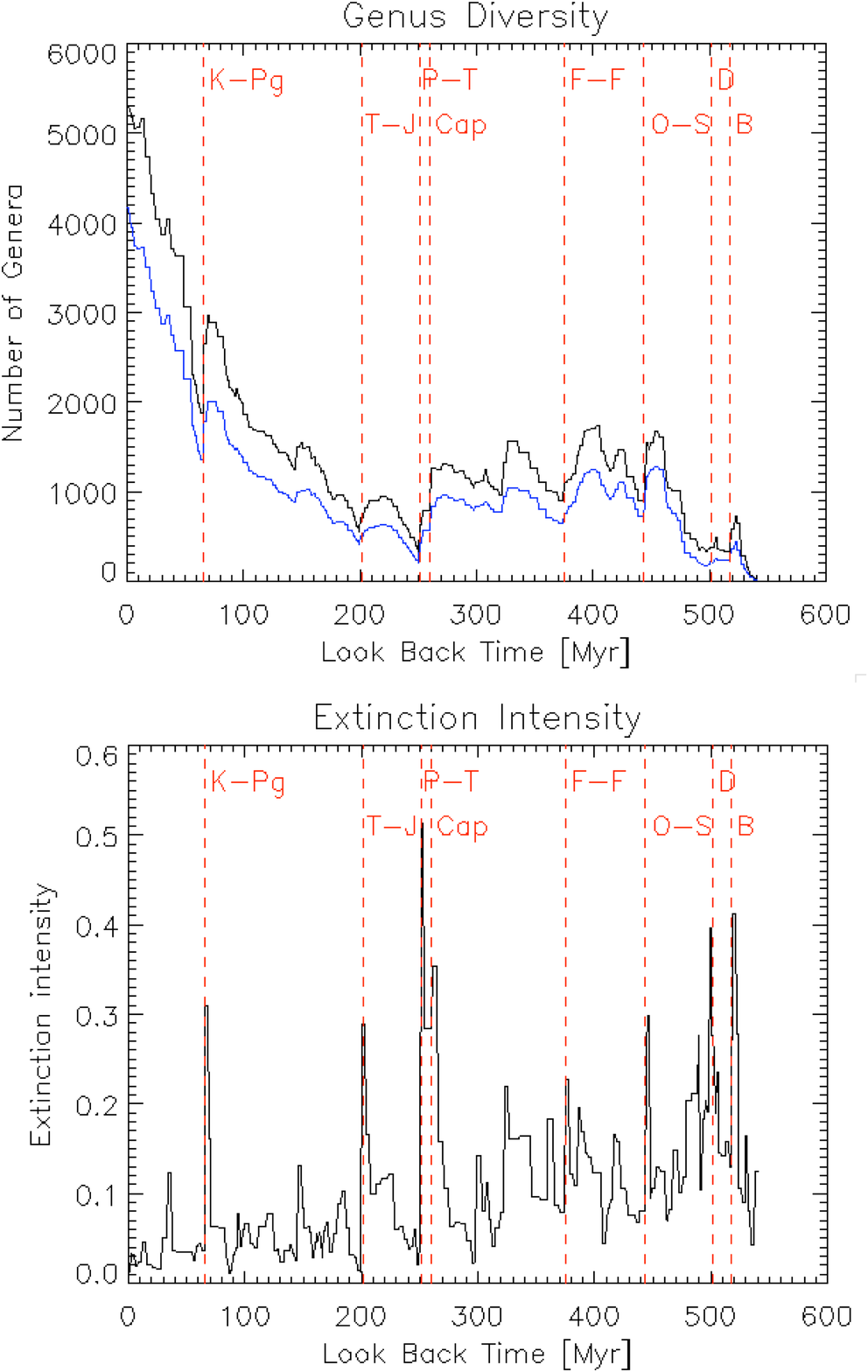}
\caption{Biodiversity in the marine fossil record. 
Top: The number of known marine animal genera as a function of time for all data (black) and with single occurrence and poorly dated genera removed (blue). 
Bottom: Extinction intensity as a function of time. 
The six major mass extinctions (O-S, F-F, Cap, P-T, T-J, and K-Pg) and two more big extinctions (B and D) are visible. 
The data used in these figures are from reference \cite{2005Natur.434..208R}.}
\label{fig:biod}
\end{figure}

Based on the Sepkoski's compendium \cite{Sepkoski2002}, 
a biodiversity database has been created from the Phanerozoic Eon fossil record \cite{2005Natur.434..208R},
and the study described in this paper is based on these data.
Figure \ref{fig:biod} (top) shows the number of known marine animal genera as a function of time for all data (black),
and data with single occurrence and poorly dated genera removed (blue). 
The six major mass extinction events\cite{10.1130/2019.2542(14),Rampino2019}, the Ordovician-Silurian extinction at 443.8 Myr ago (O-S), 
the Late Devonian extinction at 372.2 Myr ago (F-F), 
the Capitanian extinction at 259.8 Myr ago (Cap), the Permian-Triassic extinction at 251.9 Myr ago (P-T), 
the Tiassic-Jurassic extinction at 201.4 Myr ago (T-J), and the Cretaceous-Paleogene extinction at 66 Myr ago (K-Pg), 
are clearly seen in Figure \ref{fig:biod} (top). 
Five of the six major mass extinctions were most likely related to flood-basalt volcanism, and one (K-Pg) to the massive impact of an asteroid \cite{10.1130/2019.2542(14)}.
Figure \ref{fig:biod} (bottom) shows the extinction intensity as a function of time.
Extinction intensity is defined as the fraction of well-resolved genera (those having both a distinct first and last appearance known to the stage level) 
present in the bin that are absent in the following bin. 
Two more big extinction events at around 500 Myr ago, the End Botomian extinction (B) at 517 Myr ago and the Dresbachian extinction (D) at 502 Myr ago, 
are also visible in Figure \ref{fig:biod} (bottom). 
Although the details of these two extinctions are unclear due to the paucity of fossil records at that time,  
these data have also been analyzed without arbitrarily dismissing them in this work.
Further details about these data are described by reference \cite{2005Natur.434..208R}.

\begin{figure}[tb]
\centering
\includegraphics[scale=0.55]{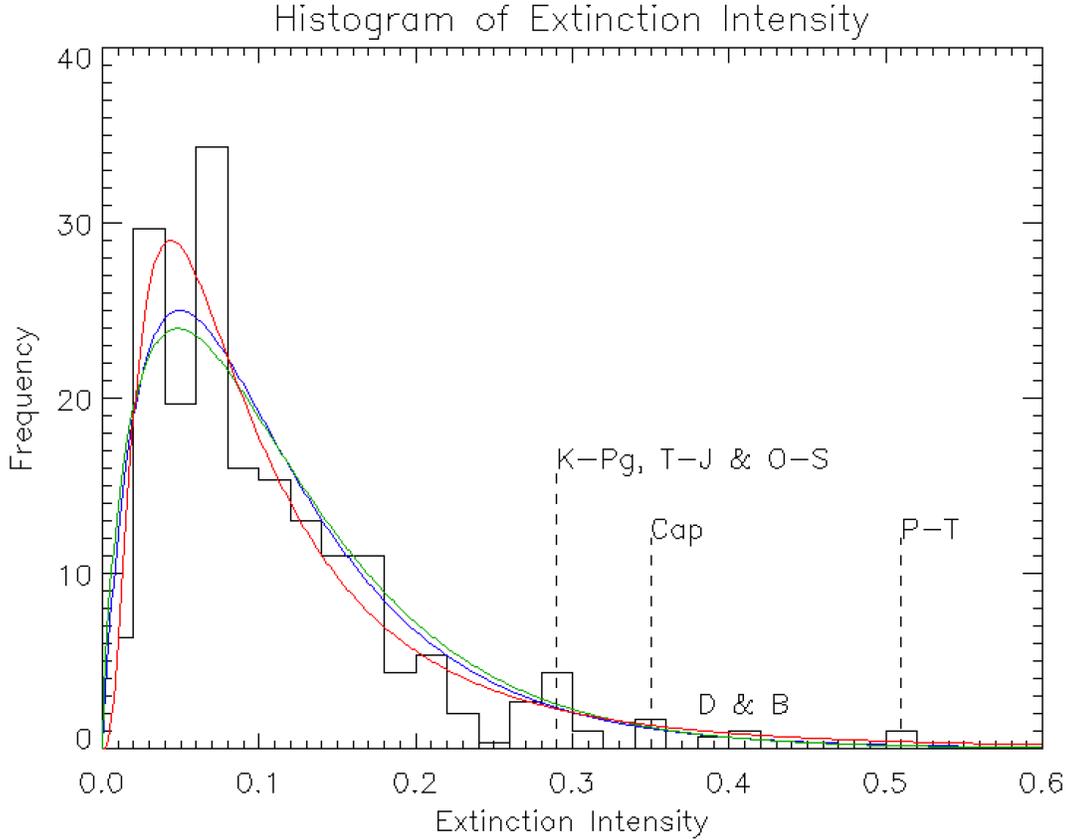}
\caption{Histogram of extinction intensity. 
The lines show the best-fitting curves for a log-normal distribution function (red), beta prime distribution function (blue), and gamma distribution function (green).}
\label{fig:hist}
\end{figure}

A histogram of extinction intensity, constructed using the extinction intensity history data shown in Figure \ref{fig:biod} (bottom), is shown in Figure \ref{fig:hist}.
Although the provided data \cite{2005Natur.434..208R} have a time bin size of 1 Myr, the time resolution of these data is closer to $\sim 3$ Myr because the peaks of the big extinctions are three bins wide. 
Although extinctions are most likely sudden events, the extinction peaks are spread out because the fossil record is incomplete (the Signor-Lipps effect \cite{10.1130/SPE190-p291}). 
Therefore, the frequency of the histogram (vertical axis in Figure \ref{fig:hist}) was divided by three to match the time resolution of 3 Myr.

This histogram was then fitted with a log-normal distribution function $\varphi_{ln}(x)$:
\begin{equation}
\varphi_{ln}(x)=\frac{1}{\sqrt{2\pi} \sigma x} \exp \left(-\frac{(\ln x - \mu )^2}{2\sigma^2}\right)
\end{equation}
where $x$ denotes the extinction intensity as a random variable, and $\mu$ and $\sigma$ are free parameters in this distribution function. 
Since the histogram has a peak at $x \sim 0.05$, minor mass extinctions ($x < 0.2$), in addition to massive extinctions, affect the overall shape of this histogram.
The choice of the log-normal distribution function is supported by the statistical principle that a random multiplicative process converges to a log-normal distribution owing to the central limit theorem.
Since the terrestrial extinctions are caused by random events, such as 
volcanic activities \cite{Hesselbo2002, KAMO200375, 10.1130/G38940.1}, asteroid impacts \cite{Schulte1214}, 
superflares of the Sun \cite{azurc3464, 2017ApJ...848...41L}, gamma-ray bursts \cite{2004IJAsB...3...55M}, and so on,
it is justified that the terrestrial extinction intensity can be expressed by a log-normal probability distribution as a result of these random multiplicative processes.
The ultimate use of a log-normal distribution has precedence in the context of the Drake equation \cite{2010AcAau..67.1366M, MACCONE201163, 2019AcAau.155..118B}.

\begin{figure}[H]
\centering
\includegraphics[scale=0.65]{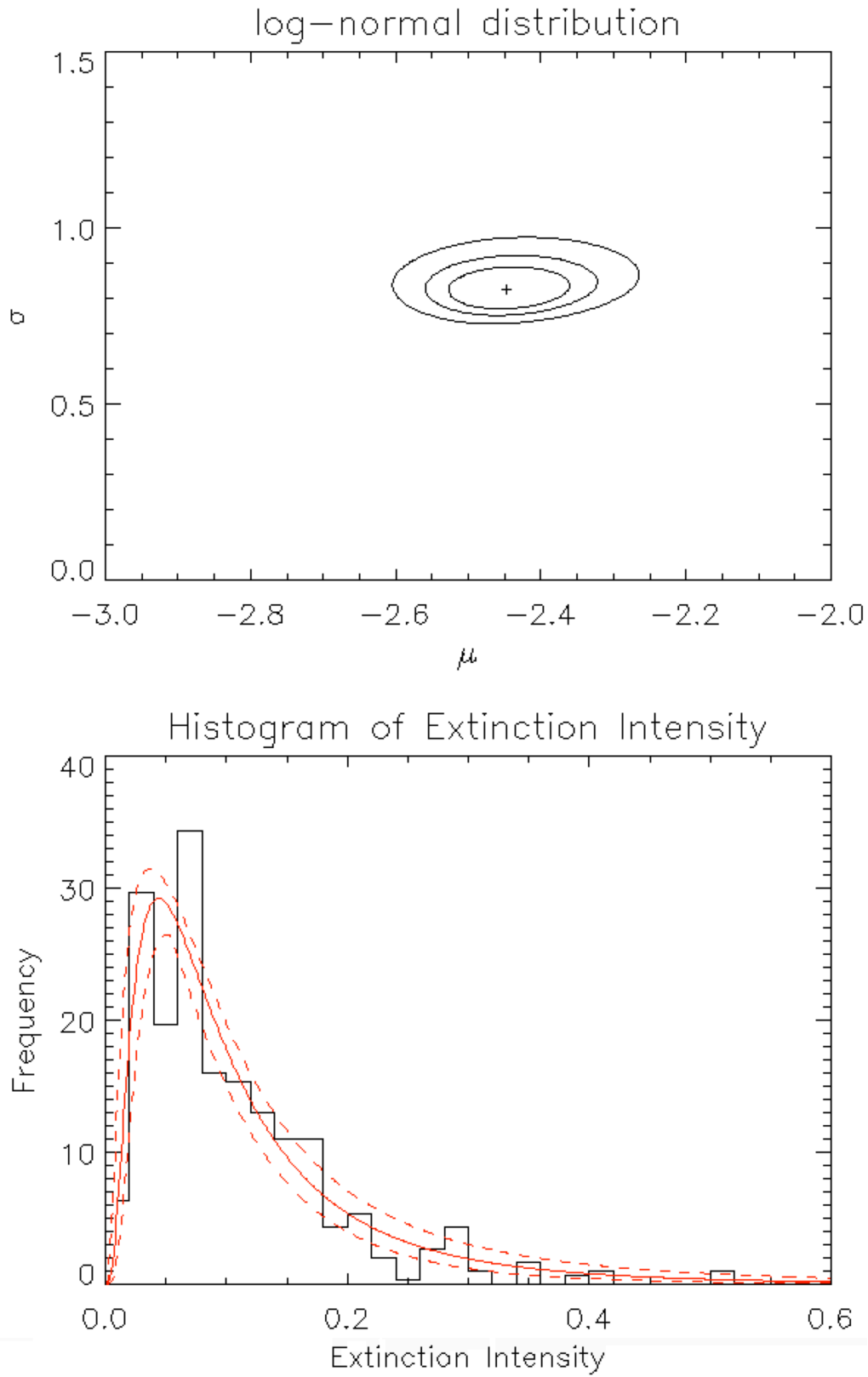}
\caption{Fitting the histogram of extinction intensity with a log-normal distribution function. 
Top: Contour map of $\chi ^2$ of the fitting. 
The cross mark at the center of the contour shows the best-fitting parameter set, and the three contour curves show 68\%, 90\%, and 99\% confidence levels, respectively. 
Bottom: Histogram of the extinction intensity and its best-fitting log-normal distribution function (red curve).
The dashed red curves constrain the 99\% confidence level region.}
\label{fig:fit}
\end{figure}

The best-fitting parameters of the log-normal function fitted to the histogram are $\mu = -2.447$ and $\sigma = 0.825$ with a scaling factor,
and its reduced chi-squared ($\chi ^2$) is 0.988. 
The best-fitting curve is shown in Figure \ref{fig:hist} and Figure \ref{fig:fit} (bottom) as a red line. 
Figure \ref{fig:fit} (top) shows the confidence contour maps of the fitted parameters. 
The uncertainties associated with this fitting were evaluated as follows. 
First, all parameter sets within a 99\% confidence level in the confidence contour map (Figure \ref{fig:fit} top) were extracted, 
and then envelops of the log-normal distribution functions with these extracted parameters are shown by dashed lines in Figure \ref{fig:fit} (bottom). 
Therefore, the two envelope curves in Figure \ref{fig:fit} (bottom) denote the uncertainties of this fitting with 99\% confidence level.

The same fitting procedures were applied to the histogram data using two additional distribution functions:
the beta prime distribution function $\varphi_{\beta}(x)$ and the gamma distribution function $\varphi_{\gamma}(x)$:
\begin{equation}
\varphi_{\beta} (x) = \frac{1}{B(\alpha, \beta)}\frac{x^{\alpha-1}}{(1+x)^{\alpha+\beta}}
\end{equation}
\begin{equation}
\varphi_{\gamma}(x) = \frac{\lambda^k}{\Gamma(k)}x^{k-1}e^{-\lambda x}
\end{equation}
where $B(\alpha, \beta)$ denotes the beta function with two free parameters of $\alpha$ and $\beta$, 
and $\Gamma(k)$ denotes the gamma function with two free parameters of $k$ and $\lambda$. 
These functions were previously used to estimate the factors in the Drake equation \cite{2019AcAau.155..118B}.
Figure \ref{fig:hist} also shows the best-fitting curves of these distribution functions. 
The reduced $\chi ^2$ of these fits for the beta prime distribution function and the gamma distribution function are 1.184 and 1.329, respectively.
These poorer fitting statistics imply that the log-normal distribution function is more appropriate for the considered histogram, which lends
credence to the assumption that the terrestrial extinction process is based on a random multiplicative process.
Therefore, only the fitting results from the log-normal distribution function are considered hereafter.

\section*{Estimation of the survival probability of terrestrial life}

The previous section showed that the histogram of extinction intensity in the Phanerozoic Eon could be well fitted by a log-normal distribution function (Figure \ref{fig:hist} and \ref{fig:fit}). 
In this section, a model is proposed to estimate the probability that life on Earth has not gone extinct from its birth to the present epoch.
In this model, the histogram of extinction intensity is interpreted as a probability distribution, representing how many fractions of all genera become extinct within a unit timescale (3 Myr),
if an extinction event of a certain magnitude, $x$, as a random variable with this probability distribution occurs every 3 Myr.
Although $x$ was defined originally as the fraction of extinct genera, i.e., $0 \le x \le 1$, here we interpret $x$ as the magnitude of each extinction event,
where $x=1$ means an extinction event with a minimal magnitude for all life on Earth to become extinct, and $x > 1$ means extinction events with greater magnitudes than it.
In this model, to illustrate the extinction history, this ``extinction lottery'' is drawn once every 3 Myr.
If the result of this lottery is $x = 0.05$, it means that 5\% of the genera on Earth became extinct during this period, 
and then the next lottery is drawn in the next 3 Myr.
If the extinction intensity $x$ takes a value of 1 or greater, it means all life on Earth are extinct, and the game is over. 
The fact that we are here now means that life on Earth has endured repeated extinction lotteries every 3 Myr since the birth of life (i.e. $x < 1$ for all lotteries). 
Under such a condition, the probability that life on Earth has survived from the origin of life to the present, $f_{i, \oplus}$, is calculated below.

The cumulative distribution function, $\Phi_{ln}(x)$, of the log-normal distribution function, $\varphi_{ln}(x)$, can be expressed as:
\begin{equation}
\Phi_{ln}(x) = \int_{0}^{x}\varphi_{ln}(x^{\prime})dx^{\prime} = \frac{1}{2}\mathrm{erfc}\left(-\frac{\ln x-\mu}{\sqrt2 \sigma} \right)
\end{equation}
where $\mathrm{erfc}(x)$ denotes the complementary error function. 
A value of $p = \Phi_{ln}(1)$ means the probability that the extinction intensity, $x$, as a random variable takes a value smaller than 1.
This can be taken to mean that not all life on Earth becomes extinct; in other words, some genera on Earth have survived through a unit timescale (3 Myr).
The history of evolution on Earth shows that life is quite resilient, because it eventually recovers even after big extinction events \cite{Chen2012}. 
Therefore, life persist unless the result of the lottery is $x>1$.
Since ``winning an extinction lottery'' is defined as the result of $x<1$, the probability of winning this extinction lottery is $p$.
Thus, the survival probability $f_{i, \oplus}$ for duration $T$ can be expressed as a probability of winning $T / \Delta T$ times in the repeated extinction lotteries,
\begin{equation}
f_{i, \oplus}(T) =  p^{T / \Delta T} \label{eq:fi}
\end{equation}
where $\Delta T$ is the time resolution of the probability distribution function ($\Delta T = 3$ Myr in this case). 

Using the best-fitting parameter set of $\mu$ and $\sigma$, the value of $p$ is calculated as $p = \Phi_{ln}(1) = 0.9985^{+0.0012}_{-0.0058}$ (99\% confidence level),
which means that the probability that some genera on Earth survive for 3 Myr is $\sim$99.85\%,
or the probability that all life on Earth becomes extinct during a 3 Myr period is $\sim$0.15\%.
Therefore, the estimated survival probability of life on Earth during the Phanerozoic Eon ($T = 540$ Myr) is 
$f_{i, \oplus}(540\  \textrm{Myr}) = 0.76^{+0.01}_{-0.06}$. 
This means that life on Earth had a $\sim$24\% probability of becoming extinct during the Phanerozoic Eon.
This value is quite reliable because the log-normal distribution function was obtained by fitting to the histogram of the extinction history in the Phanerozoic Eon.

Recent geological evidence has suggested that life on Earth first occurred 3.7-4.1 Gyr ago \cite{1996Natur.384...55M, 1999Sci...283..674R, 2002Natur.418..627V, 2014NatGe...7...25O, 2015PNAS..11214518B, 2018AsBio..18..343P}. 
Assuming that the value of $p$, obtained from the fossil record during the Phanerozoic Eon (540 Myr ago to present), 
can be extended to the entire history of life on Earth (3.7-4.1 Gyr ago to present), the survival probability for the entire history of life can be calculated as 
$f_{i, \oplus}(3.7\  \textrm{Gyr}) = 0.16^{+0.01}_{-0.03}$ or $f_{i, \oplus}(4.1\  \textrm{Gyr}) = 0.13^{+0.01}_{-0.03}$.
Therefore, as a conclusion, the probability that life on Earth survived without becoming completely extinct can be estimated as $\sim$15\% based on the fossil records of extinction from the Phanerozoic Eon.

\section*{Evaluation of the model assumptions}
Although the extinction intensity, $x$, is defined as the fraction of extinct genera, i.e. $0 < x < 1$, 
the histogram of $x$ was fitted by the log-normal distribution function $\varphi_{ln}(x)$ defined in $0 < x < \infty$.
Therefore, one might think it is better to use a truncated log-normal distribution function defined in $0 < x < 1$, $\varphi_{ln}^{\prime}(x)$:
\begin{equation}
\varphi_{ln}^{\prime}(x) = \frac{1}{\Phi_{ln}(1)-\Phi_{ln}(0)}\varphi_{ln}(x) \label{eq:tln}
\end{equation}
The fitting procedure was conducted with the truncated log-normal distribution function and the same best-fitting parameters of $\mu = -2.447$ and $\sigma = 0.825$ are obtained.
Because the difference between $\varphi_{ln}^{\prime}(x)$ and $\varphi_{ln}(x)$ is only a scaling factor with a given parameter set of $\mu$ and $\sigma$, 
and the scaling factor is determined by fitting the function to the data,
it is mathematically correct that the same best-fitting parameters are obtained with the log-normal distribution function $\varphi_{ln}(x)$ 
and the truncated log-normal distribution function $\varphi_{ln}^{\prime}(x)$.
This result gains more credibility to employ the log-normal distribution function.

The obtained value of $f_{i, \oplus}$ represents the probability that life on Earth has survived various random extinction events since the birth of life to the present.
One big assumption in this model is that the obtained extinction rate determined over only the last 540 Myr (the Phanerozoic Eon) 
can be extended to the entire history of life on Earth ($\sim$4 Gyr); however there is no guarantee that this assumption is correct.
For example, Earth experienced the Late Heavy Bombardment (LHB) 3.8-3.9 Gyr ago, which likely destroyed almost all life present on Earth at that time \cite{Cohen1754, Nisbe2001, Line2002, Zahnle2007},
but this truly massive extinction event was not included in the model.
In addition, even within the Phanerozoic Eon, the extinction rate declines with time \cite{macleod2013great}, which can be seen in Figure \ref{fig:biod} (bottom).
Moreover, the dataset used in this model (Figure \ref{fig:biod}) was constructed using fossil records of marine animal genera, not all types of life \cite{2005Natur.434..208R}.
Therefore, it is not clear whether the modern extinction rate of marine animal genera, which is modeled in this paper, can be applied to all types of life across all of history of Earth.
In this paper, however, we assumed that the modern extinction rate of marine animal genera could be applied to all life throughout history.
This is a big assumption, but this is the best that can be made at this point with the available data.

The purpose of the Drake equation is to deal specifically with questions of if, how, when and how often evolution leads to complex life,
which cannot be answered completely without other examples of evolution.
Therefore, by using the terrestrial history as a template for the histories of life on exoplanets, 
we have attempted to provide some useful perspective using the available information.
In this context, it was assumed that the obtained survival probability, $f_{i, \oplus}$, can be used to represent $f_{i}$ in the Drake equation, i.e., $f_i = f_{i, \oplus} \sim 0.15$,
because the only available data pertain to the history of Earth.
This assumption means that the evolutionary history of Earth is universal, i.e., 
once the origin of life is accomplished, the evolution of complex life always take place in any stable, sufficiently extensive environment if it does not become extinct first \cite{life6030025}.
This assumption is called the ``{\it Planet of the Apes}'' hypothesis \cite{Lineweaver2008} or the astrobiological Copernican principle \cite{Westby2020}.
A unique point of the method used here was to model extinction, rather than the appearance of life, to address the Drake equation.

\section*{Application to other estimations}

According to Equation (\ref{eq:fi}), the survival probability of life on Earth, $f_{i, \oplus}$, has two parameters:
$p$ (the survival probability for a time-bin of $\Delta T$ = 3 Myr) and $T$ (the evolution duration from the birth of life to the emergence of intelligent life).
Assuming that values for Earth, $p = 0.9985$ and $T \sim$ 4 Gyr, are universal, $f_i = f_{i, \oplus} \sim 0.15$ was obtained.
This expression of $f_i$ has a room for adjusting these two parameters for other life-bearing exoplanets to match their local environments and situations.
For example, some exoplanets have harsher environments than Earth, which would require any life to ensure stronger stellar winds or interstellar radiation fields than those in our Solar system;
hence we can readjust the survival probability so that $p$ has a smaller value. 
If the evolution speed of life on some exoplanets is slower than on Earth, we can apparently readjust the evolution time duration so that $T$ is larger. 
It is still difficult to estimate $p$ and $T$ for other exoplanets, 
but it can provide some useful constraints of $f_i$ based on scientific observations of exoplanets.

This approach can also be extended to the Seager equation, a parallel version of the Drake equation \cite{2018IJAsB..17..294S}. 
The Seager equation estimates the number of planets with detectable signs of life by way of biosignature gases as:
\begin{equation}
N^{\prime} = N_{\ast} \cdot f_{Q} \cdot f_{HZ} \cdot f_{O} \cdot f_{L} \cdot f_{S}
\end{equation}
\begin{quote}
\begin{description}
  \item[$N^{\prime}$:] The number of planets with detectable signs of life by way of biosignature gases.
  \item[$N_{\ast}$:] The number of stars in the survey.
  \item[$f_{Q}$:] The fraction of stars in the survey that are suitable for planet finding (e.g., quiet non-variable stars or non-binary stars).
  \item[$f_{HZ}$:] The fraction of stars with rocky planets in the habitable zone.
  \item[$f_{O}$:] The fraction of those planets that can be observed, according to limitations of planet orbital geometry or other limiting factors.
  \item[$f_{L}$:] The fraction of planets that have life.
  \item[$f_{S}$:] The fraction of planets with life that produce a detectable biosignature gas by way of a spectroscopic signature.
\end{description}
\end{quote}
The model introduced in this study can be applied to estimate $f_S$ in the Seager equation. 
Here, we consider molecular oxygen (O$_2$) as the favored biosignature gas, which is a gas produced by life that can accumulate to detectable levels in an exoplanetary atmosphere. 
The permanent rise to measurable concentrations of O$_2$ in the atmosphere of Earth via photosynthesis of prokaryotic and eukaryotic organisms in the ocean, 
known as the Great Oxidation Event (GOE), occurred around 2.4 Gyr ago \cite{SESSIONS2009R567, 2014Natur.506..307L}. 
Therefore, Earth took 1.3-1.7 Gyr from the birth of life to be detectable by astronomical spectroscopic observations of biosignature gas by outside observers. 
The probability that life on Earth survives until the GOE, $f_{S, \oplus}$, can be calculated using the same method shown in Equation (\ref{eq:fi}), which yields 
$f_{S, \oplus}(1.3\ \textrm{Gyr}) = 0.52^{+0.01}_{-0.06}$ and $f_{S, \oplus}(1.7\ \textrm{Gyr}) = 0.42^{+0.01}_{-0.06}$.
Assuming again that the time taken for photosynthesis to evolve is the same as that on Earth,
these values can be interpreted as $f_S$ in the Seager equation.
A value of $f_S = 0.5$ was originally speculated \cite{2018IJAsB..17..294S}, which is a reasonable estimate.

This model also can be applied to estimate the probability that existing life, including human beings on Earth, would become extinct before Earth became inhabitable.
As the Sun brightens due to its natural evolutionary process, Earth will become uninhabitable in the far future due to rising temperatures.
According to one model, the complete loss of oceans of Earth may occur in little over 2 Gyr from present, thereby transforming Earth into a desert planet.
This suggests that most forms of life will unable to survive for more than 1.3 Gyr from the present day on Earth \cite{doi:10.1002/2015JD023302}. 
The survival probability of a 1.3-Gyr period in our model is $f_{i, \oplus}(1.3\ \textrm{Gyr}) \sim 0.5$, 
indicating that existing life on Earth, including humans, have a $\sim$50\% probability of becoming extinct before Earth becomes inhabitable.
This survival probability is much higher than the estimated total longevity of our species of 0.2 million to 8 million years at the 95\% confidence level under the Copernican principle \cite{Gott1993}.
This difference may occur because advanced civilizations would be affected by much smaller catastrophes 
that would not show up as mass extinctions in the geological records. 
For example, asteroid impactors with 1-km diameters (making a 20-km diameter crater) are expected to occur about every $10^5$ years \cite{Chapman1994}
and large volcanic eruptions that could cause ``volcanic winter'' are expected to occur about every $5 \times 10^4$ year \cite{RAMPINO2002562}; 
both types of events are capable of destroying or greatly affecting an advanced civilization.
Such s discussion, however, is more related to the factors $f_c$ and $L$ in the Drake equation, rather than $f_i$ estimated in this paper, which are critical parameters to be considered
when searching for intelligent civilizations.

\section*{Conclusions} 

A new approach to estimating the survival probability of life on Earth since its birth, $f_{i, \oplus}$, was introduced.
The principle idea is that the extinction history of Earth, based on the marine fossil record, can be used to obtain the survival probability of life since it began on Earth.
The obtained value is $f_{i, \oplus} \sim 0.15$.
Under the astrobiological Copernican principle \cite{Westby2020}, this survival probability can be interpreted as $f_i$ in the Drake equation, i.e., $f_i = f_{i, \oplus} \sim 0.15$.
Because $f_{i, \oplus}$ is a two-parameter function of $p$ (the survival probability for a unit time) and $T$ (the evolution time from the birth of life to intelligent life), 
this method can be extended to estimate the survival probability on other life-bearing exoplanets by adjusting these two parameters to the local environments of the considered exoplanets.

\bibliography{Tsumura}

\section*{Acknowledgements }
The author would like to thank Yasuhisa Nakajima (Tokyo City University) for discussions regarding paleontology,
and Enago (www.enago.jp) for the English language review.
This research was supported by JSPS KAKENHI Grant Number 18KK0089 and 20H04744.

\section*{Author contribution}
K.T. performed all work in this paper.

\section*{Competing interests}
The author declares no competing interests.

\end{document}